\documentclass[aps, reprint, floatfix, amsmath, amssymb]{revtex4-2}
\usepackage{float} 
\usepackage{graphicx}
\usepackage{dcolumn}
\usepackage{bm}
\usepackage{xcolor}
\begin{document}

\preprint{APS/123-QED}

\title{Construction and analysis of surface phase diagrams to describe segregation and dissolution behavior of Al and Ca in Mg alloys}

\author{Jing Yang}
\affiliation{Department for Computational Materials Science, Max-Planck-Institut für Eisenforschung GmbH, Max-Planck-Str. 1, D-40237 Düsseldorf, Germany}

\author{K. B. Sravan Kumar}
\affiliation{Department for Computational Materials Science, Max-Planck-Institut für Eisenforschung GmbH, Max-Planck-Str. 1, D-40237 Düsseldorf, Germany}

\author{Mira Todorova}
\email{m.todorova@mpie.de}
\affiliation{Department for Computational Materials Science, Max-Planck-Institut für Eisenforschung GmbH, Max-Planck-Str. 1, D-40237 Düsseldorf, Germany}
 
\author{Jörg Neugebauer}%
\affiliation{Department for Computational Materials Science, Max-Planck-Institut für Eisenforschung GmbH, Max-Planck-Str. 1, D-40237 Düsseldorf, Germany}

\date{\today}

\begin{abstract}
Segregation and dissolution behavior of Mg alloyed with Ca and Al are studied by  performing density functional theory calculations considering an extensive set of surface structures and compositions. Combining ab initio surface science approaches with cluster expansion for ordered surface structures we construct surface phase diagrams for these alloys. We utilize these diagrams to study segregation phenomena and chemical trends for surfaces in contact with a dry environment or with an aqueous electrolyte. We show that the presence of water dramatically impacts the stability and chemical composition of the considered metallic surfaces. We furthermore find that the two alloying elements behave qualitatively different: whereas Ca strongly segregates to the surface and becomes dissolved upon exposure of the surface to water, Al shows an anti-segregation behavior, i.e., it remains in Mg bulk. These findings provide an explanation for the experimentally observed increase/decrease in corrosion rates when alloying Mg with Al/Ca.
\end{abstract}

\maketitle

\section{\label{sec:Intoduction} Introduction} 

Magnesium with its very low density of 1.74\,g/cm$^3$~\cite{Kittel} and a price comparable to that of commonly used aluminum alloys shows great promise towards developing lightweight materials in automotive and aerospace industries. Mg and its alloys have also great potential for improving the energy density in batteries as either cathode and anode materials~\cite{VanNoorden2014}, due to the ability of Mg to carry twice as much charge as Li and its lower cost. Other promising applications relate to their use as biodegradable implants, by exploiting the non-toxic nature of Mg, its mechanical properties similar to bones and its ability to corrode easily, which could circumvent the need for implant removal~\cite{Witte2008}. 

Despite having applications in such diverse fields, Mg based alloys suffer from two major limitations: poor ductility~\cite{Sandlobes2017} and poor corrosion resistance~\cite{Yang2016}. The ductility of Mg can be significantly improved upon alloying with small amounts of the inexpensive and non-toxic Al (1\,wt.\%) and Ca (0.1\,wt.\%)~\cite{Sandlobes2017}. At higher concentrations of a few weight percent, the same alloying elements (Al and Ca) refine the grain size of Mg, thereby improving its mechanical strength~\cite{Lee2000}. The reduction of the Mg grain sizes is due to the inhibition of the grain boundary motion during re-crystallization by finely dispersed Mg$_{2}$Ca particles forming at calcium contents of 2-3\,wt.\,\%, as shown by high-ratio differential speed rolling (HRDSR)~\cite{Seong2015}. Overall, alloying of Mg with Al and Ca improves its mechanical properties. 

Alloying with Al and Ca has also an impact on the corrosion properties of Mg. Increasing the Ca content in Mg alloys leads to an increased anodic corrosion potential~\cite{Kirkland2011}. This is consistent with the observation that Mg$_{2}$Ca has a larger anodic corrosion potential compared to pure Mg indicating that Ca is anodic in Mg alloys~\cite{Sudholz2011}. Anodic corrosion potentials imply that metal atoms become oxidized and dissolve  in the electrolyte as positively charged metal ions (Mg$^{2+}$, Ca$^{2+}$).  The corrosion rate becomes significantly higher for Ca concentrations above the solubility limit in Mg~\cite{Kirkland2010, Nabilla2016, Gusieva2015}. In contrast, Mg alloys with 2-10\,wt.\,\% Al, as well as Mg-Al intermetallics (Mg$_{2}$Al$_{3}$, Mg$_{17}$Al$_{12}$), show an increased cathodic potential compared to Mg, i.e. they are less prone to corrosion~\cite{Sudholz2011, Leslie2017}. Indeed, the corrosion rates of Mg alloys were found to decrease upon increasing Al content~\cite{Guangling2003, Gusieva2015}. The anodic nature of Ca in the corrosion experiments points towards a preferential oxidation of Ca, which acts as a sacrificial anode upon exposure to water. In contrast, Al with its cathodic nature is expected to remain in a Mg alloy. 

This picture is also confirmed by computations. Electrochemical potentials calculated using a Born-Haber cycle indicate that the Mg-Al intermetallic Mg$_{17}$Al$_{12}$ is more cathodic than Mg, while the Mg-Ca intermetallic Mg$_{2}$Ca is more anodic than Mg~\cite{Thekkepat2021}. Surface energies of low index Mg surfaces with Al and Ca substituting Mg atoms in the surface layer calculated using density functional theory (DFT) indicate a higher dissolution rate for Ca  compared to Al~\cite{Nezafati2016}. Similar to the observations in the corrosion experiments, dissolution potential differences, calculated using chemical potentials derived from Mg vacancy formation energies in the presence of Al and Ca substituted in Mg surfaces, show Al to be more cathodic in nature than Ca~\cite{Nezafati2016}. While these computational investigations provide a first qualitative understanding of the impact Al and Ca have on Mg corrosion~\cite{Thekkepat2021, Nezafati2016} critical questions remain. For example, the surface structures of the Mg alloys, dependencies on the alloy concentration and the impact of water on the surface segregation behavior are less explored. 

A step in this direction is provided by a DFT study looking at the segregation behaviour of various alloying elements substituting a Mg atom in the first layer for the single coverage of $\frac{1}{9}$\,ML (monolayer; at 1\,ML all atoms in the topmost Mg layer would be substituted by an alloying element)~\cite{Wang2021}. This study showed that segregation of Ca from the bulk to the Mg\,(0001) surface is thermodynamically favored. However, the systematic exploration of the dependence of Al and Ca surface segregation on alloy concentration, as well as the transfer of this information to construct surface phase diagrams, is lacking. 

In this study, we therefore employ DFT calculations and focus on the initial, i.e. non-oxidised state of Mg alloys with dilute bulk concentrations of Al and Ca, constructing surface phase and surface Pourbaix diagrams. To identify the stable surface structures, we consider two scenarios: i) the surface of as cast Mg-Al and Mg-Ca alloys in the absence of an oxidizing medium (in vacuum) and ii) the surface when the as cast alloy is exposed to a corrosive environment. Accounting for the relevance of water for wet corrosion we include it in the calculations for the later case, by modelling it as an implicit solvent~\cite{Mathew2014, Mathew2019}. The dependence of the surface segregation behavior on surface coverage is accounted for by considering different surface coverages ranging from as low as $\theta=\frac{1}{36}$\,ML up to a full monolayer. We limit this study to the Mg(0001) surface, which is the most stable close-packed hcp surface termination~\cite{Vitos1998} and has been widely studied both computationally and experimentally~\cite{Namba1981,Bungaro1997,Schroder2004,Kim2005,Pozzo2009}. Though realistic surface structures are more complex, low index surfaces such as the Mg(0001) serve as a model system to study generic behavior and chemical trends. 

\section{\label{sec:methods} Methods} 
\subsection{\label{sec:methods-DFT} Density functional theory calculations}
All DFT calculations are performed in a plane wave based DFT framework and the projector augmented wave (PAW) approach as implemented in the Vienna Ab-initio Simulation Package (VASP)~\cite{Kresse1993, Kresse1996}. Exchange and correlation are described with the PBE implementation of the Generalized Gradient Approximation (GGA)~\cite{Perdew1996}. We employ Fermi smearing of 0.1\,eV, converge energies to $10^{-5}$\,eV and optimize geometries using a force criterion of 0.01\,eV/{\AA}. Based on careful convergence checks we chose an energy cutoff of 360\,eV for further calculations, as it ensures that the lattice constants and the cohesive and formation energies of all considered materials are accurate to at least $5 \times 10^{-3}$\,{\AA} and 5\,meV each. Surfaces are modelled using six layer thick symmetric slabs and a vacuum region of 12\,{\AA}, removing the need to apply a dipole correction~\cite{Neugebauer1992} to the electrostatic potential in $z$-direction. For the Mg$(0001)$ ($1 \times 1$) surface cell we use a ($12 \times 12 \times 1$) $k$-point mesh, which is equivalently folded for larger cells. This set-up ensures that mixing energies for Ca and Al incorporation in the Mg$(0001)$ surface are converged to better than 10\,meV per Ca or Al atom. To study the impact of water on the considered surfaces, the vacuum region in the supercell is replaced by a solvent using an implicit solvation model based on Poisson-Boltzmann as implemented in VASPsol~\cite{Mathew2014, Mathew2019}. For these calculations we used eight layer thick symmetric slabs to obtain work functions and interface energies.

\subsection{Calculating surface energies }
A key step in computing the segregation behavior of alloying elements on the Mg surface is to identify which surface structure is the thermodynamically most stable one for a given set of thermodynamic state variables. Since surfaces are thermodynamically open systems - they can exchange atoms with the bulk or with the environment to which the surface is exposed - the relevant state variables are temperature and chemical potentials of the chemical species forming the alloy. For the binary system considered here these are Mg alloyed with Al or Ca, i.e., $\mu_{\rm Mg}$, $\mu_{\rm Al}$, and $\mu_{\rm Ca}$. The impact of pressure is small~\cite{Vandewalle2002,Reuter2001} and will be neglected in this study. The relevant energy is the surface formation energy $E^f_\sigma(T, \mu_{\rm Mg}, \mu_{\rm X})$ with X either Al or Ca. The subscript $\sigma$ runs over all considered surface configurations. The one which minimizes the surface formation energy for a given set of state variables is the thermodynamically stable phase
\begin{equation}\label{eq:sigma_eq}
  E^{\rm surf}_{\sigma_{\rm eq}}(T, \{\mu_{i}\}) = \min_{\sigma} E^{\rm surf}_{\sigma}(T, \{\mu_{i}\}), 
\end{equation}
where the index $i$ runs over Mg and the alloying element X.

We note that the surface formation energy $E^{\rm surf}_\sigma(T,\{\mu_i\})$ is nothing else than the Grand Potential $\Theta_\sigma(T, \{\mu_i\})$, which is routinely and very successfully used to compute bulk phase diagrams, e.g. in the Monte Carlo (MC) based ATAT code~\cite{ATAT} or defect phase diagrams, e.g. using MC/MD in LAMMPS~\cite{LAMMPS, Pan2017}. We will use in this article the term surface formation energy, since this is the commonly used name in the surface science community where this formalism and terminology are in use for almost three decades~\cite{Qian1988, Northrup1993, Neugebauer1996,Felice1996, Neugebauer1998, Chen2000, Vandewalle2002-2, Vandewalle2002, Neugebauer2003, Wang2000, Reuter2001}. We also prefer this notation since it conveys in the context of surfaces a clear physical interpretation: The surface formation energy is the excess energy required to create the surface from ideal bulk.

Following the surface science approach~\cite{Neugebauer1996, Neugebauer1998, Vandewalle2002}, the above introduced surface formation energy is defined as:
\begin{eqnarray}
\label{eq:surf_energy_general_expression}
\begin{split}
    E^{\rm surf}_{\sigma}(T, \{\mu_{i}\}) & = \frac{1}{2A}[E_{\sigma}^{\rm DFT} - TS_{\rm surf}
    \\ & - \sum_{i = \rm Mg, X}n_{i,\sigma}  (E_{i-\rm bulk}^{\rm DFT} + \mu_{i})  
     ] .
    \end{split}
\end{eqnarray}

Here, $E_{\sigma}^{\rm DFT}$ is the DFT computed total energy of the surface structure $\sigma$. $n_{\rm i,\sigma}$ represents the number of atoms of species $i$ in the surface structure $\sigma$.  $A$ is the area of the surface unit cell described by the respective supercell and the factor two takes into account that the symmetric slabs we consider here consist of two equivalent surfaces. 
$S_{\rm surf}$ is the excess surface entropy, which is dominated by configurational entropy. Further contributions include electronic or vibrational entropy, which are substantially smaller and will therefore be neglected in this work. For the ordered phases, the configurational entropy is zero, i.e., $S_{\rm surf}=0$. It should be noted that by neglecting the entropy term, we are treating the ordered surface phases in the same way as line compounds in bulk surface diagrams. However, while a bulk line compound exists, as indicated by its name, only for a single bulk concentration, an ordered surface phase with zero configurational entropy can exist over a finite range of bulk alloying composition (see Fig. \ref{fig:T_conc_Diag}a).

The surface structures are characterized by the local arrangement of the solute atoms in the first surface layers. Since, as will be shown later, solute energies already in the second layer are very close to the ones in Mg bulk we consider only surface structures with solutes in the first two layers. The solute concentration of a surface configuration $\sigma$ is described by the coverage $\theta_\sigma$, which is defined by the number of solute atoms in this layer divided by its total number of atoms.

\subsection{\label{sec: chemical_potentials} Chemical potentials}
A critical input for calculating the surface energies according to Eq.~\ref{eq:surf_energy_general_expression} are the chemical potentials. These potentials provide a high-level link between thermodynamic concepts and specific experimental condition or control parameters such as bulk composition, temperature, ion concentration in the case of wet corrosion etc. In this section we discuss their definition, how to obtain them from \textit{ab initio} calculations, their physical and thermodynamic boundaries as well as how to translate experimental conditions into chemical potential and vice versa.

\subsubsection{Introduction and references}

The chemical potentials $\mu_{\rm Mg}$ and $\mu_{\rm X}$ describe the chemical reservoirs with which the Mg and X atoms are in equilibrium. Rather than using absolute chemical potentials as obtained when using the DFT total energies directly we follow common convention and reference our potentials with respect to a physically well-defined configuration.

A major advantage of using a physics based reference is that the chemical potentials do not contain spurious pseudopotential and code specific contributions but just the code-agnostic energy difference between reference and actual value.
The potential consists then of two parts (see Eq. \ref{eq:surf_energy_general_expression}): The DFT reference energy and the chemical potential $\mu_i$ with respect to this reference. For the Mg-X alloy system natural reference systems are Mg hcp bulk (for Mg) and bulk fcc Al/Ca (for Al/Ca). Thus, the chemical potential for Mg bulk is $\mu_{\rm Mg\text{-}bulk}=0$, independent of the chosen DFT code or pseudopotential. This choice also implies that the bulk contribution in the slab region is completely eliminated (see Supplemental Material (SM), Sec. A~\cite{SM}). As a consequence, the surface energy defined by Eq.~\ref{eq:surf_energy_general_expression} is by construction the excess energy necessary to create a surface from bulk. 
Since we focus here on substitutional surface
structures, i.e. structures where a surface Mg atom is substituted by an Al or Ca atom, formally a Mg/X atom has to be brought/taken into/out of the thermodynamic reservoir $\mu_{\rm Mg}$/$\mu_{\rm X}$ for Mg/X atoms, respectively. Thus, only the difference between
the chemical potentials of the host and solute species enters in Eq.~\ref{eq:surf_energy_general_expression}, i.e., $E^{\rm surf}_{\sigma}(T, \mu_{\rm Mg} - \mu_{\rm X})$.

\begin{figure}[b]
\centering
\includegraphics[width=0.40\textwidth]{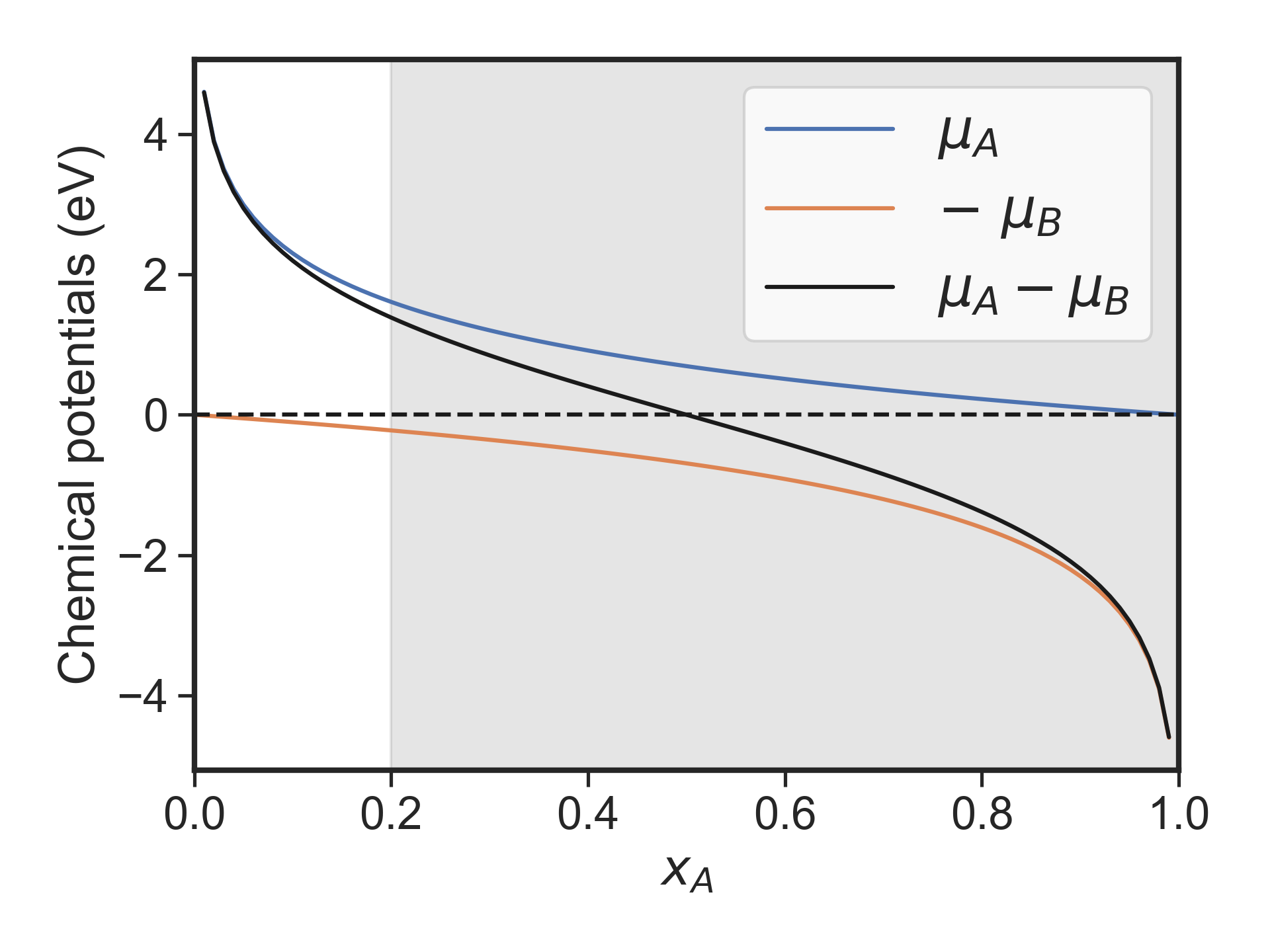}
\caption{\label{fig:ChemicalPotential} Difference between the chemical potentials of species A and B in an A$_{\rm x_A}$B$_{1-{\rm x_A}}$ compound. Also shown are the chemical potentials of the two species. Note that we plot $-\mu_{\rm B}$ rather than $\mu_{\rm B}$ to show the matching asymptotic behavior with $\mu_{\rm A} - \mu_{\rm B}$ for $\rm x_A$ approaching 1. }
\end{figure}

Fig.~\ref{fig:ChemicalPotential} shows the chemical potential difference between host and solute chemical potential, as well as the individual host and solute chemical potentials. The latter are given by $\mu=-k_{\rm B} T \ln x$ with $x$ the concentration of the host/solute in the bulk. As can be seen, up to a solute concentration of $\approx 10$\,\% the difference in chemical potentials can be safely replaced by the chemical potential of the solute alone, i.e.,  
\begin{equation}
    (\mu_{\rm X} - \mu_{\rm Mg}) \approx \mu_{\rm X} \quad ,
\end{equation}
\begin{equation}
    \mu_{\rm Mg} \approx 0.
\end{equation}
The latter equation simply means that the Mg reservoir is that of Mg bulk in the dilute solute limit. Since our focus will be on solid solutions and precipitation of Al and Ca occurring at bulk concentrations at about 1\,at.\% the difference can be safely replaced by the chemical potential of the alloying element alone.

\subsubsection{\label{sec:chemical_potentials_boundaries}The chemical reservoirs }
In a realistic scenario, the chemical reservoirs with which a surface is in contact are the bulk on one side and the environment on the other. Since the total number of solute atoms on the surface is negligibly small compared to the total number of solute atoms in the bulk or environment, the surface chemical potential (see SM, Sec. B~\cite{SM}) will equilibrate towards one of these embedding chemical potentials through an exchange of solute atoms.

\begin{figure}
\centering
\includegraphics[width=0.48\textwidth]{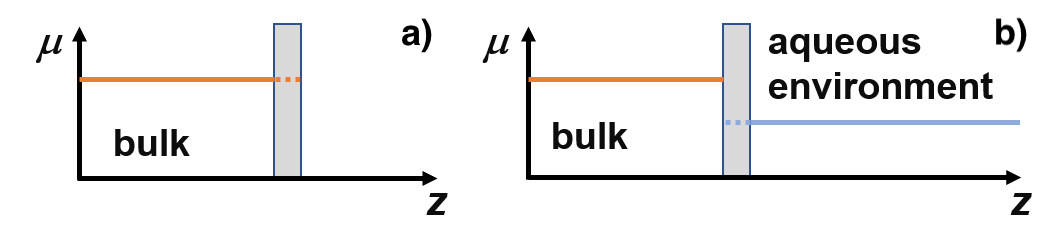}
\caption{\label{fig:schematic} Chemical potential of the alloying element along the surface normal for the two scenarios considered in the paper: (a) The surface is in contact with a dry environment (vacuum). For this case the surface is assumed to be in thermodynamic equilibrium with the bulk. (b) The surface is in contact with an aqueous electrolyte. Since adsorption/diffusion kinetics between surface and electrolyte is much faster than diffusion between surface and bulk the surface is assumed to be in thermodynamic equilibrium with the electrolyte.}
\end{figure}

In the following, we will discuss two scenarios (see Fig.~\ref{fig:schematic}): First, the case when the surface is equilibrated with the bulk alloy composition, i.e., after casting the alloy and before exposing it to an oxygen environment. The second scenario addresses the case of wet corrosion, in which the surface comes into contact with a water electrolyte. In this case, solutes are constantly exchanged between the surface and their solvated state as ions in solution.

In the dilute limit, i.e. in the absence of solute-solute interaction, the chemical potential of a solute atom X in Mg bulk is
\begin{equation}\label{eq:solute_chem_pot}
    \mu_{\rm X\text{-}in\text{-} Mg\text{-}bulk}(x_{\rm X}, T) = \Delta E_{\rm X\text{-}in\text{-} Mg\text{-}bulk} + k_B T \ln x_{\rm X} 
\end{equation} 
with 
\begin{equation}
    \Delta E_{\rm X\text{-}in\text{-}Mg\text{-}bulk} = E^{N-1}_{\mathrm{Mg} + 1 \mathrm{X}}-\frac{N-1}{N}E^{N}_{\mathrm{Mg}}-E_{\rm X\text{-}bulk}.
\end{equation} 
Here $E^N_{\mathrm{Mg}}$ represents the DFT total energy of Mg bulk with $N$ atoms, $E^{N-1}_{\mathrm{Mg}+ 1 \mathrm{X}}$ is the energy of the same bulk system with one Mg atom substituted by the alloying atom X. $E_{\rm X\text{-}bulk}$ is the bulk reference of X. With this definition and at T=0\,K the chemical potential $\mu_{\rm X\text{-}in\text{-}Mg\text{-}bulk}$ becomes $\Delta E_{\rm X\text{-}in\text{-}Mg\text{-}bulk}$=119\,meV for Al and 95\,meV for Ca. The positive sign here implies that bringing an Al/Ca solute from its native bulk (i.e. Al or Ca bulk) into Mg is an endothermic reaction at $T=0$ K. The solute concentration is then solely driven by the configurational entropy (second term in Eq. \ref{eq:solute_chem_pot}).

If the surface is in thermodynamic equilibrium with its bulk, the surface chemical potential entering Eq.~\ref{eq:surf_energy_general_expression} becomes $\mu_{\rm X}= \mu_{\rm X\text{-}in\text{-}Mg\text{-}bulk}(x_X, T)$, i.e., a given experimental scenario can be easily translated into a corresponding solute chemical potential. An obvious  advantage of using the chemical potential is that it combines two independent variables (concentration and temperature) in a single one. This allows for an intuitive $xy$-representation of surface energies vs chemical potential in the absence of an explicit temperature dependence. This is the case for ordered surface structures where $S_{\rm surf}=0$ in Eq.~\ref{eq:surf_energy_general_expression}. Since the chemical potential is a rather abstract quantity, we select and discuss in the result section a few representative scenarios. For the Mg-X bulk alloys,  we consider bulk mole fractions of $2\times 10^{-2}$ and $10^{-3}$ at 800 K for Al and Ca respectively, mirroring experimentally synthesized solid solutions in these alloys. 

When the surface is exposed to an aqueous electrolyte containing ions, we consider the chemical potentials for Al$^{3+}$ and Ca$^{2+}$ ions (Fig.~\ref{fig:schematic}b). For Ca, these values are calculated at the Mg dissolution potential at which Mg dissolves as Mg$^{2+}$. The dissolution potential $U=-2.51$\,V is calculated from the bulk Mg Pourbaix diagram for a Mg$^{2+}$ concentration of 10$^{-5}$\,mol/L, representative of the Mg$^{2+}$ concentrations observed in experiments~\cite{Nowak}. Using the dissolution potential of Mg ($U^{\rm Mg} = -$2.51 V) for Al, the fcc Al bulk becomes the stable phase rather than the ionic species Al$^{3+}$ in solution. As a consequence, Al$^{3+}$ solutes will precipitate on the surface, giving rise to an Al-water interface. The Pourbaix diagram of Al shows however a much lower voltage $U^{\rm Al}=-$1.676 V at which dissolution starts~\cite{CRC}. We therefore choose this voltage to determine the chemical potential of Al ions in solution. For Ca, the dissolution potential is very high ($U^{\rm Ca}= -$2.868 V~\cite{CRC}), making Ca$^{2+}$ ion the preferred species at the Mg dissolution potential. For this case we therefore use the dissolution potential of Mg.

The dependence of chemical potentials of Al and Ca on their ion concentrations~\cite{Todorova2014} are given by:

\begin{equation}\label{eq:Al_ion_chem_pot}
\mu_{\rm Al} = \Delta G^{0}_{{\rm Al}^{3+}} + k_{\rm B}T \cdot ln\frac{c_{{\rm Al}^{3+}}}{c_{0}} + 3e(U_{\rm SHE} - U^{\rm Al}),
\end{equation}

\begin{equation}\label{eq:Ca_ion_chem_pot}
\mu_{\rm Ca} = \Delta G^{0}_{{\rm Ca}^{2+}} + k_{\rm B}T \cdot ln\frac{c_{{\rm Ca}^{2+}}}{c_{0}} + 2e(U_{\rm SHE} - U^{\rm Mg}),  
\end{equation}
where $\Delta G^{0}_{{\rm Al}^{3+}}$ and $\Delta G^{0}_{{\rm Ca}^{2+}}$ are the formation energies of Al$^{3+}$ and Ca$^{2+}$ ions respectively. For the standard hydrogen electrode potential ($U_{\rm SHE}$) we chose $-$4.73 V, similar to previous studies~\cite{Todorova2014}, and consider concentrations of $c_{\rm Al^{3+}} = 3\times 10^{-7}$ and $c_{{\rm Ca}^{2+}} = 8\times 10^{-8}$\,mol/L, representative of concentrations observed in experiments~\cite{Nowak}. $c_0$ is the reference concentration, which is equal to 55.55 mol/l (1 l of water contains 55.55 mol of
H$_2$O molecules). Additional details of the derivation of Eq.~\ref{eq:Al_ion_chem_pot} and Eq.~\ref{eq:Ca_ion_chem_pot}  can be found in SM, Sec. C~\cite{SM}.

\subsubsection{\label{sec:surf_phase_diag_seg_method}Thermodynamic boundaries of the chemical potentials}

The chemical potential of element X in its thermodynamically stable intermetallic phases is generally lower than that of its bulk phase. Therefore, they serve as upper bounds for the chemical potential $\mu_{\rm X}$. For Al, the Mg containing intermetallic phase where Al has its lowest chemical potential is Al$_{12}$Mg$_{17}$. For Ca it is Mg$_2$Ca. The respective chemical bulk potentials are $\mu_{\rm Ca}(\mathrm{Mg_2Ca})=-0.23$\,eV and $\mu_{\rm Al}(\mathrm{Al_{12}Mg_{17}})=-0.09$\,eV (see SM, Sec. D~\cite{SM}). For chemical potentials $\mu_{\rm X}$ larger than these potentials the solid solution becomes thermodynamically unstable against the formation of Al$_{12}$Mg$_{17}$ or Mg$_2$Ca precipitates, respectively. This upper bound of the solute chemical potential allows us to compute the thermodynamic solubility limit of element X in Mg bulk as a function of temperature following Eq. \ref{eq:solute_chem_pot}.

When kinetic barriers prevent the system from reaching its thermodynamically preferred configurations, it is possible to realize chemical potentials well above the critical solute potential $\mu_{\rm X}$ at which the intermetallic precipitates would form in thermodynamic equilibrium. Experimentally this can be achieved by methods such as rapid quenching (see SM, Sec. E~\cite{SM}). In these cases, there still exists an upper limit of $\mu$. As evident from Eq.~\ref{eq:solute_chem_pot}, the upper limit of the (equilibrium) solute chemical potential in bulk Mg is $\mu_{\rm X\text{-}in\text{-}Mg\text{-}bulk}(x_{\rm X}, T) < \Delta E_{\rm X\text{-}in\text{-}Mg\text{-}bulk}$. In the following discussions, we will therefore discuss both the equilibrium solubility limit and the $\Delta E_{\rm X\text{-}in\text{-}Mg\text{-}bulk}$ limit for supersaturated solid solution.


\subsection{Cluster expansion}

\begin{figure}[t] 
\centering
\includegraphics[width=0.4\textwidth]{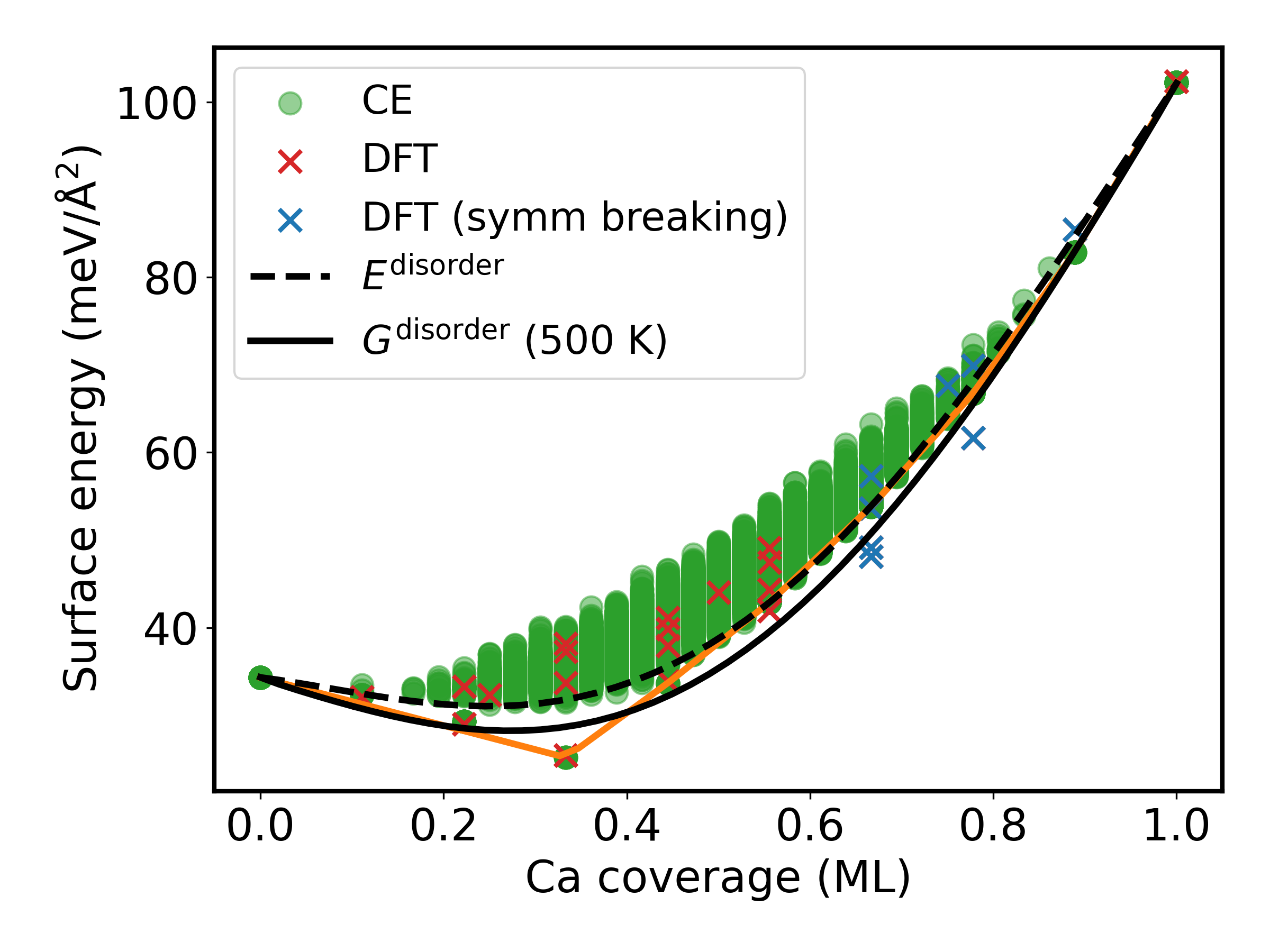}
\caption{\label{fig:CE} Cluster expansion (CE) prediction of the surface energies (green) upon substitution of Ca in the Mg(0001) surface. The CE calculations are done on a (6$\times$6) surface cell with 10$^5$ random Ca distributions. 
The orange line shows the corresponding convex hull. The crosses show the energies of ordered phases on (2$\times$2) and (3$\times$3) surfaces obtained with DFT calculations. The structures which undergo symmetry breaking (see text) are marked with blue. The CE predictions corresponding to the DFT-calculated structures are marked by large green dots. From the CE predictions, we obtain $E^{\rm disorder}$ by fitting a parabola that envelopes the energies of the disordered phase (dashed black line).   }
\end{figure}

To systematically explore the large configuration space of possible surface structures $\sigma$, we combine DFT calculations with the cluster expansion (CE) approach. The CE parameters are obtained by fitting to DFT calculations going over all possible symmetry-inequivalent configurations of the alloying atoms on the (2$\times$2) and (3$\times$3) Mg(0001) surface. The CE includes pair and triple interactions. In Fig.~\ref{fig:CE}, we show the surface energies calculated by DFT (red crosses) and the surface energies of random structures on a (6$\times$6) surface calculated with CE (green dots). We observe that for coverages up to 0.6 ML, the CE predictions agree well with DFT results. However, for $\theta = \frac{2}{3}$ and $\theta = \frac{7}{9}$, the CE overestimates the energies of the surfaces. A careful analysis of the surface structures reveals that the surfaces undergo symmetry breaking during DFT relaxation, which cannot be accounted for in the CE (see SM, Sec. F~\cite{SM}). Nevertheless, as we will see later, such high coverages occur only at solute concentration well above the solubility limit and thus are thermodynamically unstable.

In Fig.~\ref{fig:CE}, we can clearly observe an energy gap between the lowest-energy ordered phases (red crosses on the energy-hull line) and disordered surface states calculated by CE for a large set of random configurations. This energy gap is most pronounced at $\theta=\frac{1}{3}$ and also visible at $\theta=\frac{2}{9}$ ($\approx$ 0.22). The significantly lower (i.e. energetically more favorable) surface energy  of the ordered structures originates from attractive solute-solute interactions that are absent in the disordered structure. Such an energy gap between the ordered structure and the large number of disordered structures is a prerequisite for the existence of an order-disorder transition. Above a critical temperature the ordered surface state becomes thermodynamically unstable against a disordered phase.

To describe this transition in the phase diagram, we construct in the following an analytical expression for the surface energy of the disordered phase. From CE, we can obtain the energy of all possible surface configurations $E_{\sigma}^{\rm CE}$, analogous to $E_{\sigma}^{\rm DFT}$ in Eq.~\ref{eq:surf_energy_general_expression}. We then calculate the surface energies at $T=$ 0 K following:
\begin{eqnarray}
\label{eq:CE}
\begin{split}
    E^{\rm surf}_{\sigma}(T=0\mathrm{\ K}, \{\mu_{i}\}) & = \frac{1}{2A}[E_{\sigma}^{\rm CE}
    \\ & - \sum_{i = \rm Mg, X}n_{i,\sigma}  (E_{i-\rm bulk}^{\rm DFT} + \mu_{i})  
     ].
    \end{split}
\end{eqnarray}
The obtained $E^{\rm surf}_{\sigma}(T=0\mathrm{\ K}, \{\mu_{i}\})$ values are plotted in Fig.~\ref{fig:CE} (green dots). We exclude the ordered phases and fit a parabolic curve that envelops the energies of the disordered structures to represent $E_{\theta}^{\rm disorder}(T,\{\mu_i\})$ (Fig.~\ref{fig:CE}, black dashed line). At given chemical potential, the equilibrium surface coverage of the disordered phase is the one that minimizes the surface formation energy. Following Eq.~\ref{eq:sigma_eq}, we obtain this minimized energy by 

\begin{equation}\label{eq:disorder}
    E^{\rm disorder}(T,\{\mu_i\}) = \min_{\theta}[E_{\theta}^{\rm disorder}(T=0\mathrm{\ K},\{\mu_i\})-TS_{\rm surf}],
\end{equation}
where  
\begin{eqnarray}\label{eq:config_entropy}S_{\rm surf} =-k_{\rm B}\left[\theta \cdot ln \theta + (1-\theta)\cdot ln(1-\theta) \right].
\end{eqnarray}
The coverage $\theta$ at which the energy $E_{\theta}^{\rm disorder}$ reaches its minimum is the equilibrium concentration $\theta_{\rm eq}(T,\{\mu_i\})$. It should be noted that with CE, we can also obtain $\theta_{\rm eq}(T,\{\mu_i\})$ via grand canonical Monte Carlo simulation. However, in this work our focus is to build, as for the ordered surface phases, an analytical model for disordered surface phase $E^{\rm disorder}$. For the following analysis and discussions we will therefore use the equations outlined above.

\section{\label{sec:Results_discussion} Results and discussion} 
\subsection{\label{sec:surf_seg_Al_Ca_results} Segregation behavior of Al and Ca at as cast Mg surfaces}

\begin{figure}[t] 
\includegraphics[width=0.48\textwidth]{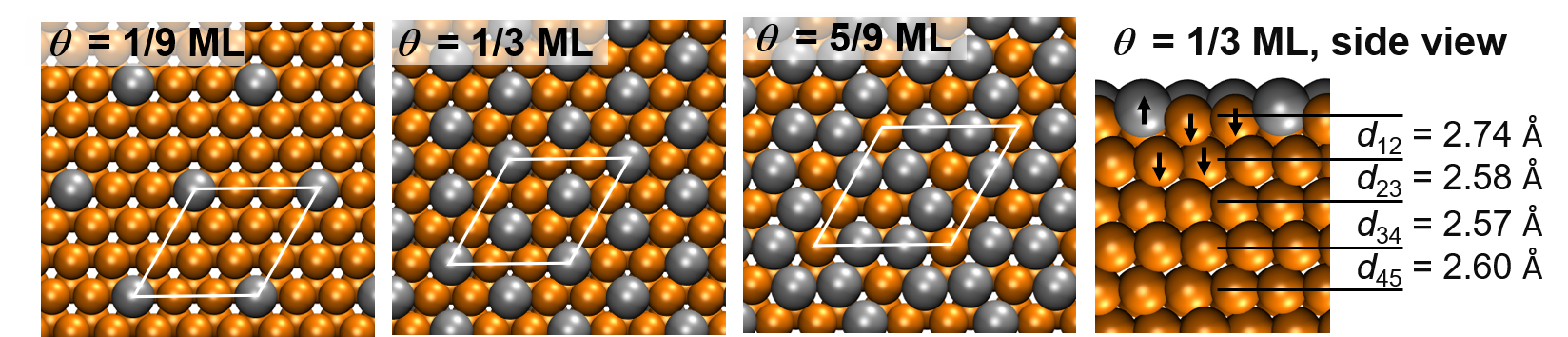}
\caption{\label{fig:Geometry} Surface structures for several coverages of Ca substituting Mg atoms in the surface layer on the Mg(0001) surface shown in top view. The lowest energy structure ($\theta_{\rm Ca} = 1/3$\,ML) is seen also in side view, showing the relaxation pattern. Mg atoms are shown as orange balls, Ca atoms as silver.}
\end{figure}

We now study the surface segregation behavior of Al and Ca in vacuum by applying the formalism outlined in Sec.~\ref{sec:methods}. We thus systematically substitute the Mg atoms in the top surface and the sub-surface layers of a Mg(0001) slab with Al and Ca atoms thus realising different surface structures and coverages. An illustration of some of the studied Ca substitutions in the surface and sub-surface layers of Mg are shown in Fig.~\ref{fig:Geometry}. 

\subsubsection{\label{sec:Al_Ca_surf_phase_diags}  Surface phase diagrams of as cast surfaces}

\begin{figure}[t] 
\centering
\includegraphics[width=0.48\textwidth]{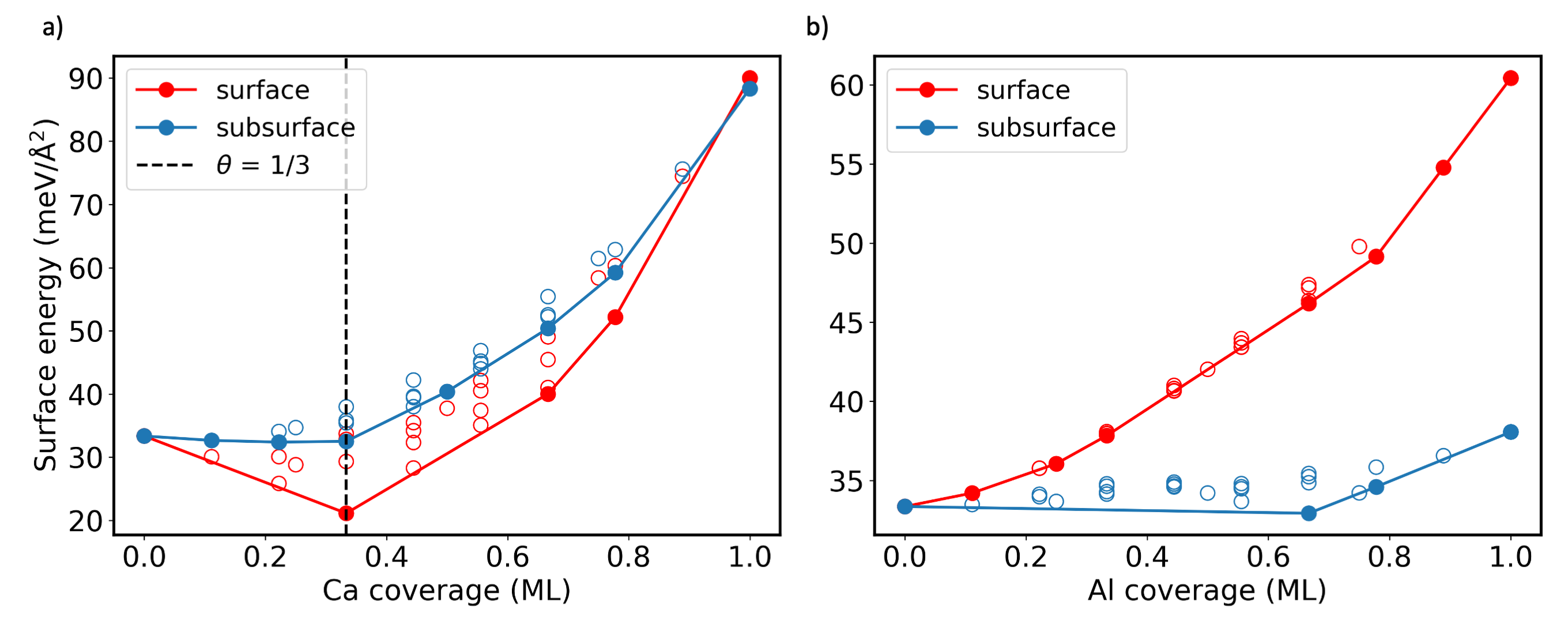}
\caption{\label{fig:ConvexHull_vac} 
Surface energies $E^{\rm surf}$ upon substitution of ({\it left}) Ca and ({\it right}) Al in the Mg(0001) surface and sub-surface layers. Surface energies are calculated in vacuum. The Al/Ca chemical potentials are that of a single Al/Ca atom in Mg bulk ($\Delta E_{\mathrm{X\text{-}in\text{-}Mg\text{-}bulk}}$). The solid lines show the convex hull construction for the top (red) and subsurface (blue) layer structures at $T = 0$\,K.}
\end{figure}

We construct surface phase diagrams for Al and Ca substitution at the Mg(0001) surface. As a first step we construct the convex hull from the DFT-computed surface energies. These energies are obtained using Eq.~\ref{eq:surf_energy_general_expression} at $T=$0 K, $\mu_{\rm Mg}=\mu_{\rm Mg\text{-}bulk}$ and $\mu_{\rm X}=\mu_{\rm X\text{-}in\text{-}Mg\text{-}bulk}$. The latter choice allows us to directly determine whether moving an X atom from the Mg bulk to the surface is energetically favorable (i.e., it reduces the surface energy by segregating to the Mg surface) or not. Fig.~\ref{fig:ConvexHull_vac} shows the convex hull construction. The phases with the lowest energy at a given $\theta$ are marked by filled dots and lie on the convex hull. We note that changing the chemical potential $\mu_{\rm X}$ will change the energy minimum of the convex hull, but not its shape (see SM, Sec. G~\cite{SM}). Comparing the energies for Ca substitution in the surface (red line) and sub-surface (blue line) layer, we see that Ca prefers the top surface except for the highest coverage $\theta=1$. Al solutes show the exact opposite trend. For both Ca and Al we find that the surface energies remain unchanged for low coverages, i.e., the energy of a solute remains virtually unchanged when bringing it from Mg bulk into the subsurface. Thus, in the low covereage regime the second layer already behaves like bulk. In the top surface layer, however, a qualitatively different behavior is observed. Up to a coverage of 0.6 ML Ca segregation leads to a substantial reduction of the surface energy, while Al increases the surface energy over the entire coverage range, even for small coverages. Thus, Ca shows strong segregation tendencies whereas Al is a strong anti-segregant.

\begin{figure}[t] 
\includegraphics[width=0.48\textwidth]{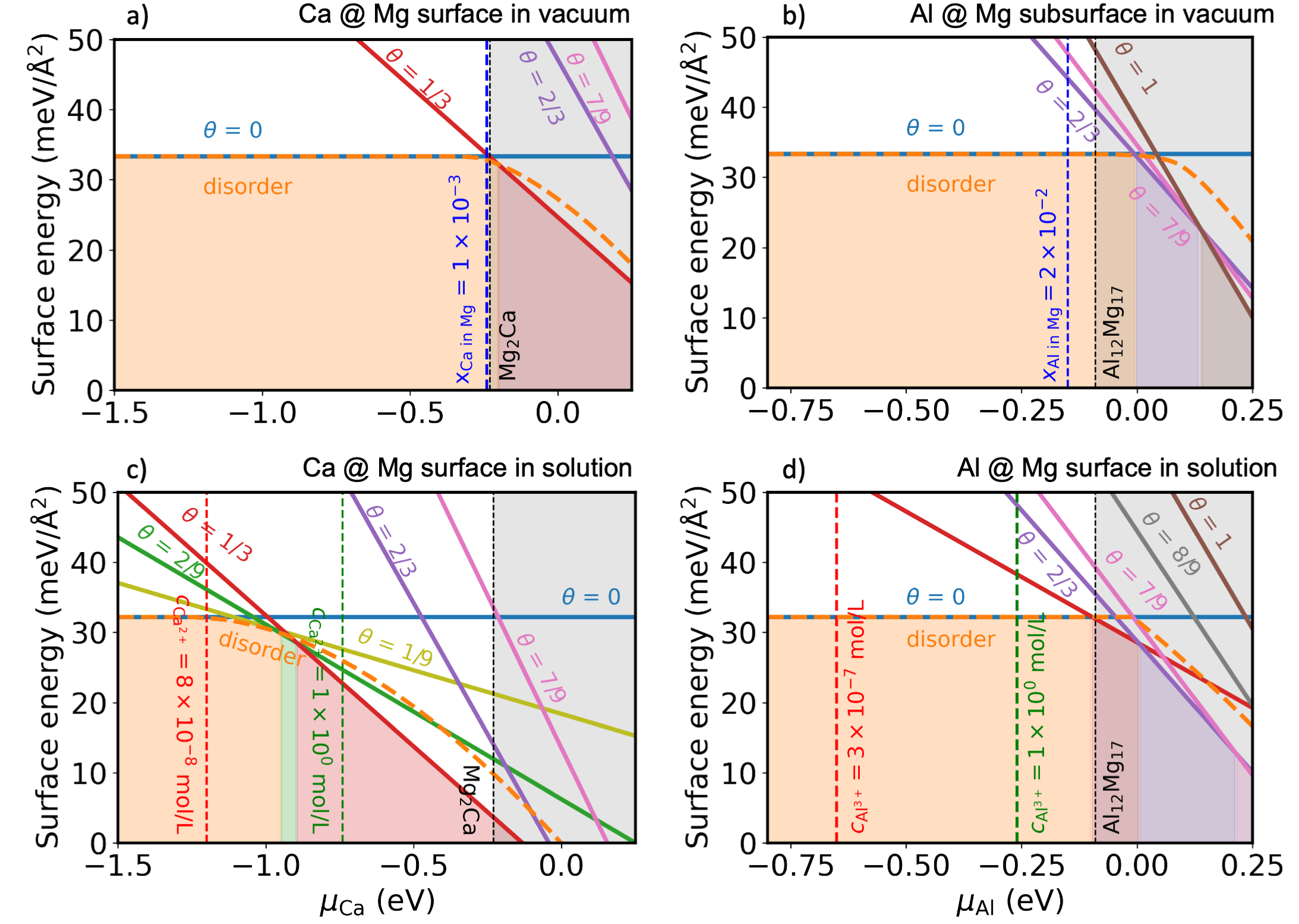}
\caption{\label{fig:SurfPhaseDiags} Surface phase diagrams for Ca and Al substitution at Mg(0001) surfaces {\it (top)} in vacuum and {\it (bottom)} in aqueous electrolyte. The surface coverages of Ca and Al are reported in ML. Chemical potentials corresponding to selected Ca/Al concentrations in Mg bulk and different Ca$^{2+}$ concentrations in solution are plotted as vertical dashed lines.  The dashed black line  corresponding to Mg$_2$Ca/Al$_{12}$Mg$_{17}$ formation marks the solubility limit of Ca and Al in hcp Mg, respectively. On the left side of this line, Ca incorporates as solid solution and on the right side of the line (grey shaded area) it is in a supersaturated state and thus thermodynamically unstable against the formation of its respective precipitates. The free energy of the disordered phase is calculated for $T=500$\,K in vacuum and $T=300$\,K in solution.} 
\end{figure}
 
Next we construct the surface phase diagrams (Fig.~\ref{fig:SurfPhaseDiags}). In these diagrams we include also the disordered surface phase (Eq.~\ref{eq:disorder}). Each colored area shows the respective surface phase that has the lowest Gibbs energy for a given range of chemical potentials. Since Al is a strong anti-segregant, the Al surface phase diagram shows only phases with sub-surface occupation.  For the surfaces in vacuum at 500 K, the disordered phase emerges as the most stable one at low Al/Ca chemical potentials, which correspond to Al/Ca poor conditions (e.g. low Al/Ca concentrations in the bulk). At low chemical potential $\mu_{\rm X}$, it is energetically unfavorable for solutes to go from bulk to the surface, resulting in low surface solute concentrations. As a consequence, the interaction between substituted solute atoms at the surface is small and configurational entropy wins over remaining (small) interactions. It should be noted that though the energies of the solute free ($\theta=0$) surfaces (blue line) appear to be identical to that of the disordered phases (orange line) at low chemical potentials in Fig. \ref{fig:SurfPhaseDiags}, the energy of the latter is lower. The energy differences of the two phases are small because the corresponding solute coverages $\theta$ are very low ($<$ 10$^{-4}$). Nevertheless, the energy of the disordered phase in the vicinity of $\theta =0$ is indeed below that of the ordered one, as shown in Fig. \ref{fig:CE}. For Ca, going to higher chemical potentials, a c$(3 \times 3)$ structure with $\theta=1/3$\,ML coverage (see Figure 4) and at even higher potentials a structure with a coverage of $\theta=2/3$\,ML become stable. We also see that already at a temperature of 500\,K, as considered here, the configurational entropy makes the disordered phase the thermodynamically favorable one at low coverages (between 0 and 1/3) as well as at higher coverages (between 1/3 and 2/3). 

Constructing the phase diagram in $\mu$ space allows us to include information from the bulk phase diagram, such as bulk phase transitions, in the same diagram. Specifically, the black dashed vertical line marks the solubility limit, i.e., the chemical potential at which the Ca/Al-Mg solid solution becomes thermodynamically unstable and starts to form Mg$_2$Ca/Al$_{12}$Mg$_{17}$ precipitates. For Ca, as shown in Fig.~\ref{fig:SurfPhaseDiags}, only the disordered surface structure (a solid solution of Ca and Mg in the top surface layer) can be realized in the solid solution region. All ordered surface structures shown in this diagram become stable only at a chemical potential well above the solubility limit and are thus thermodynamically unstable. For supersaturated Ca solid solution in Mg, which occur e.g. when rapidly quenching as-cast Mg, the Ca chemical potential can be well above the solubility limit. At these non-equilibrium conditions the ordered $c$(3$\times$3) surface structure can form. For Al, the surface phase diagram shows a similar behavior: in the thermodynamically stable region only the disordered random solution can be formed. Again, meta-stable phases may be formed when going to non-equilibrium conditions where the excess solute concentration in the bulk is well above the equilibrium limit.

\subsubsection{\label{sec:Al_Ca_surf_seg_isotherms}  Segregation isotherms for Ca at the Mg(0001) surface}

\begin{figure}[t] 
\includegraphics[width=0.48\textwidth]{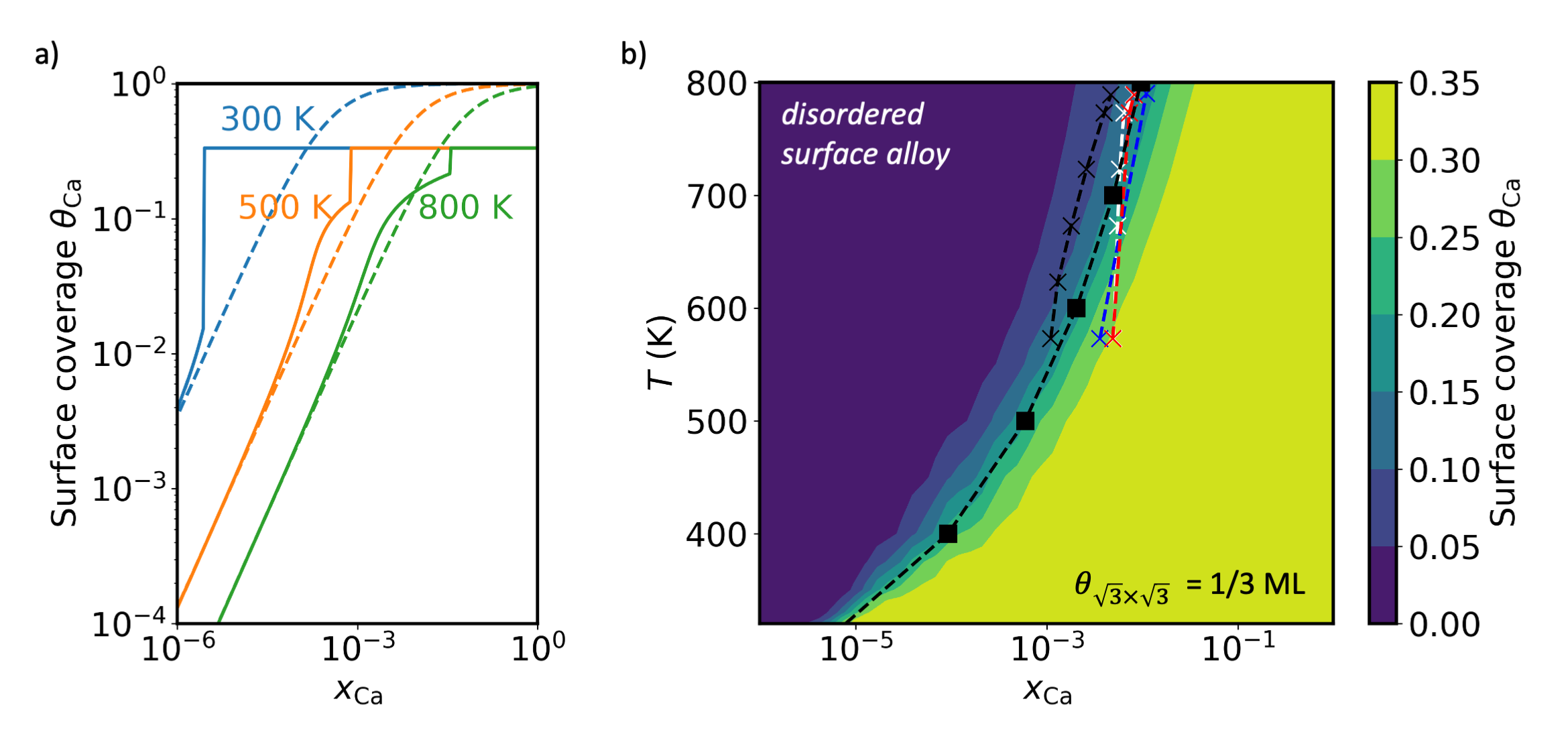}
\caption{\label{fig:T_conc_Diag} a)  Surface segregation isotherms at different temperatures for Ca segregation to the Mg(0001) surface as a function of the Ca concentration $X_{\rm Ca}$ in bulk Mg at 300, 500 and 800 K. The dashed lines show the respective prediction of the McLean model~\cite{McLean1957}. 
b) Heat map for coverages of Ca at the Mg(0001) surface at different temperatures and $X_{\rm Ca}$. Experimentally determined solubility of Ca in Mg bulk taken from Ref.~\cite{Nayeb1987}: white line~\cite{Vosskueler1937}, red line~\cite{Nowotny1940}, blue line~\cite{Haughton1937}, black line with crosses~\cite{Bulian1946}. The dashed black line with squares shows the DFT computed solubility limit calculated with Eq. \ref{eq:solute_chem_pot}.}  
\end{figure}

To connect to specific experimental scenarios the chemical potential can be expressed by physical quantities that describe the environment with which the surface is in thermodynamic contact (see Sec~\ref{sec: chemical_potentials}). In this section we will consider the initial state of as-cast surfaces, where the solute concentration at the surface and in bulk can be assumed to be in thermodynamic equilibrium (see Fig.~\ref{fig:T_conc_Diag}). In a solid solution, where the Al and/or Ca concentration is low, the corresponding  chemical potentials can be directly obtained from the bulk solute concentration and temperature via Eq.~\ref{eq:solute_chem_pot} (see Fig.~\ref{fig:ChemicalPotential}). 

In thermodynamic equilibrium the Al/Ca chemical potential in bulk and at the surface are identical. The chemical potential entering Eq.~\ref{eq:surf_energy_general_expression} to compute the surface energies can thus be substituted by Eq.~\ref{eq:solute_chem_pot}, allowing us to write it as function of temperature and bulk composition. The resulting segregation isotherms are visualized in Fig.~\ref{fig:T_conc_Diag} in two representations. In the left diagram (Fig.~\ref{fig:T_conc_Diag}\,a) the Ca surface coverage is shown as function of the Ca bulk concentration for a set of temperatures. At low Ca bulk concentrations the surface coverage increases monotonously until a phase transition to the ordered $\theta=1/3$\,ML surface structure occurs. For low coverages, the McLean isotherm~\cite{McLean1957} accurately reproduces the computed isotherms. The reason for their perfect match is that the expression for the surface energy (Eq.~\ref{eq:disorder}) becomes formally equivalent to McLean at low surface coverages (see SM, Sec. H~\cite{SM}). However, as also shown in Fig.~\ref{fig:T_conc_Diag}\,a, at high bulk solute concentrations the assumption of a single coverage independent solute defect or surface energy breaks down. As a consequence the McLean model qualitatively fails to describe this for the composition region relevant for practical applications. A remarkable difference to the continuous relation between bulk concentration and surface coverage is that the presence of the ordered structure pins the coverage, as seen in the appearance of a terrace of constant surface coverage in the plot. 

The right diagram (Fig.~\ref{fig:T_conc_Diag}\,b) shows the segregation isotherms as  a heat map depicting the Ca surface coverage at different temperatures and Ca concentrations in Mg bulk. An increase of the bulk concentration leads to an increase in surface coverage, while an increase in the temperature leads to a decrease in the surface coverage (as shown in Fig.~\ref{fig:T_conc_Diag}\,a and qualitatively explained by the McLean model~\cite{McLean1957}). As a consequence, the bulk Ca concentration required to achieve the Ca surface coverage of $\frac{1}{3}$\,ML keeps decreasing with decreasing temperature. Even at bulk Ca concentrations as low as $10^{-5}$\,at.\,\%, when going to sufficiently low temperatures ($\approx 300$\,K) this high surface coverage remains. Of course, at such low temperatures diffusion rates become exceedingly low, preventing the surface from reaching thermodynamic equilibrium/ segregation isotherms in realistic time intervals. 

\subsubsection{\label{sec:Explanation_for_surf_seg} Rationalizing the Al and Ca surface segregation behavior}

To rationalize the strong tendency of Ca to segregate to the Mg surface, as opposed to Al, we consider materials properties like the surface energies or the atomic radii of the involved elements. Looking first at the surface energies of Mg(0001) ($\gamma = 34.7$\,meV/\AA$^2$), Ca(111) ($\gamma = 34.2$\,meV/\AA$^2$~\cite{Wang2014}) and Al(111) ($\gamma = 41.8$\,meV/\AA$^2$~\cite{Marzari2009}) we note, that the surface energies of Mg(0001) and Ca(111) are very similar, while the one of Al is $16$\,\% higher. This suggests that Ca segregation to the surface of Mg will be favorable and possibly even decrease the systems surface energy, while the opposite is expected for Al. Indeed, we find that for the relevant low Ca surface coverage structures (including the $1/3$\,ML structure, which is reminiscent of a plane in the Mg$_2$Ca intermetallic) the surface energy is decreased compared to the pristine surface. For Al we observe, as expected, the opposite trend. 

The atomic radii of Mg, Ca and Al are 1.60\,{\AA}, 1.98\,{\AA} and 1.43\,{\AA}~\cite{Kittel} respectively, showing that Ca is bigger than Mg, while Al is smaller. Consequently, we expect Ca incorporation into the lattice to induce a larger compressive strain, which could be released by segregation to the surface. The release of such unfavorable strain energy would make Ca segregation to the surface energetically highly attractive. To test this hypothesis, we analyse the relaxation behavior of Ca and Al substituted in the Mg surface at different surface coverages. As expected, we find significant outward relaxations for Ca (0.5 - 0.8 {\AA}). For Al, much smaller ($\sim$ 0.2 {\AA}) and inward relaxations are observed. Thus, Ca can release a large part of the compressive strain it has in the bulk at the surface. In contrast, Al, not experiencing large strain in bulk, does not profit from this mechanism. This size-dependent surface segregation phenomenon has also been observed previously. For example, alkali atoms heavier than Li are immiscible in the bulk of Al metal due to the size mismatch~\cite{Neugebauer1992}. However, they intermix well at the surface where the strain can be relaxed~\cite{Stampfl1994,Stampfl1995,Borg2005}. 

\subsection{\label{sec:dissolution_beh_Al_Ca_results} Dissolution behavior of Al and Ca from the Mg surface}

To study the dissolution behavior of Al and Ca from the Mg(0001) surface in a corrosive environment we use again surface phase diagrams (Fig.~\ref{fig:SurfPhaseDiags}) constructed from DFT energies. These energies are obtained for coverages of Al and Ca in the top Mg(0001) surface layer. To capture the impact of the electrolyte an implicit solvent approach (VASPsol~\cite{Mathew2014}) is used. This approach emulates an aqueous environment and accounts for its impact on surface structures and energies. We freeze in the surface structures obtained in the vacuum calculations without performing additional structural relaxation. This choice is made based on a systematic study comparing surface phase diagrams obtained for surfaces in vacuum, in implicit solvent, and in explicit solvent. We found that using frozen structures of geometries relaxed in vacuum produces phase diagrams closest to the explicit solvent model. These results will be reported in a forthcoming publication.  We again employ the methodology outlined in section \ref{sec:methods}, which allows us to treat ordered and disordered structures in the same framework. The diagram (Fig.~\ref{fig:SurfPhaseDiags}) provides information not only about the surface, but also of the bulk phases.

\subsubsection{\label{sec:Ca_surf_phase_diag_solvation_impact}  Impact of solvation on stability and dissolution of the substitutional elements on Mg(0001)}

A comparison of the surface phase diagrams for Ca and Al in the absence and presence of water (Fig.~\ref{fig:SurfPhaseDiags}) reveals distinct differences. In the following, we will first focus on Ca and continue the discussion to Al. The surface phase diagram constructed for Ca substituting Mg atoms in a Mg(0001) surface in contact with water reveals a richer set of ordered phases  in the range of chemical potentials compatible with concentrations of bulk solid solution. We see the appearance of a low coverage $1/9$ and $2/9$\,ML Ca ordered phase, which is not found in the absence of water. Modelling an aqueous environment via an implicit solvent provides mainly the electrostatic screening characteristic of the solvent. Thus, the stabilisation of the $1/9$ and $2/9$\,ML structure is related to the built-up of a partial solvation shell around the Ca atom, which due to its larger size sticks out of the surface layer (see Fig.~\ref{fig:Geometry}). This solvation-induced stabilisation of Ca surface atoms is absent in the vacuum case. The solvation energy gain is apparently sufficient to overcome the impact of entropy, which dominates at low chemical potentials and favors disordered surfaces in the vacuum case. 

\begin{figure}[t]
\includegraphics[width=0.48\textwidth]{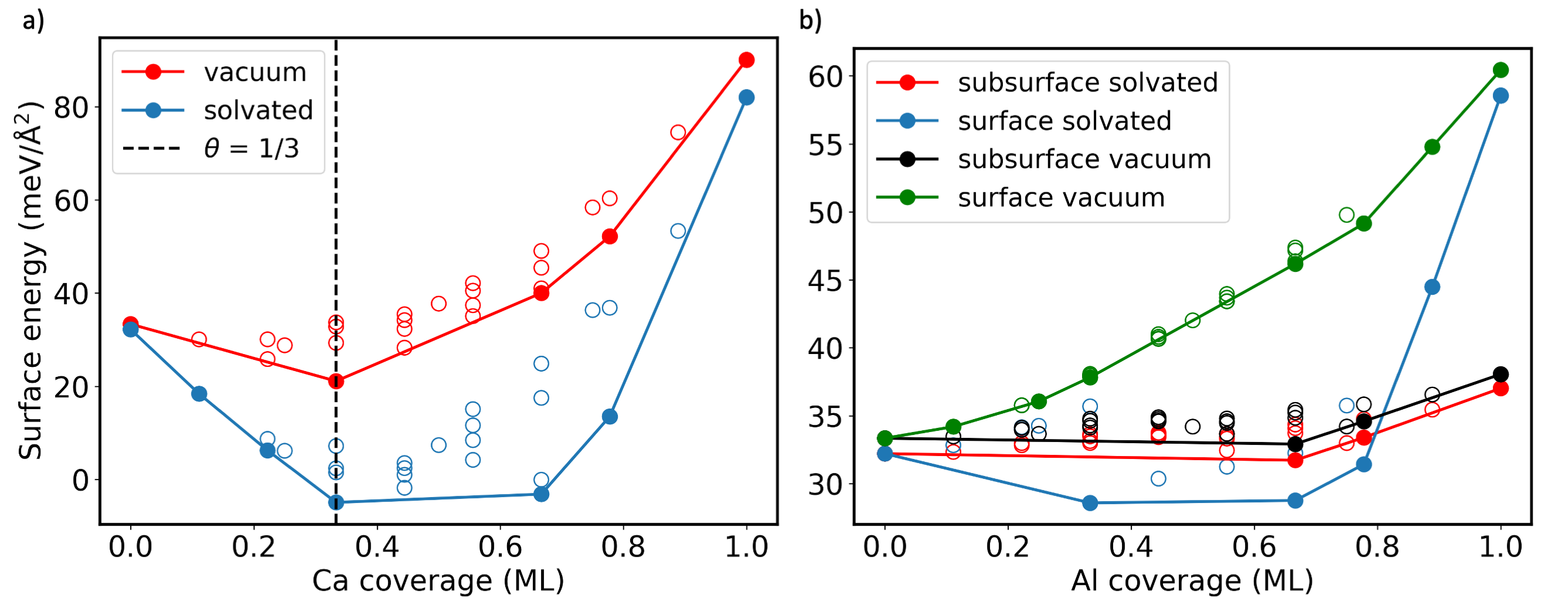}
\caption{\label{fig:ConvexHull_sol} Surface energies ($G_\textrm{surf}$) upon substitution of ({\it left}) Ca and ({\it right}) Al in the Mg(0001) surface in the absence and presence of implicit water. The Al/Ca
chemical potentials are that of a single Al/Ca atom in Mg
bulk ($\Delta E_{\mathrm{X\text{-}in\text{-}Mg\text{-}bulk}}$). For Ca, we show only results for Ca in the top surface layer 
because the structures with Ca in the subsurface layer have been found to be higher in energy (i.e. less favorable) for all coverages. For Al, results for Al residing on the subsurface layer and on the top surface are shown.}
\end{figure}

The screening of unfavorable electrostatic interactions and the ensuing gain in solvation energy, is an important stabilization mechanism for semiconductor/insulator surfaces in aqueous environment~\cite{Yoo2018}. For metals, it was considered to be of significantly lesser importance due to their large dielectric screening. The appearance of a solvation stabilized Ca structure is therefore significant, since it reveals an efficient solution induced mechanism to stabilize specific surface structures. 
Finally, the presence of the solvent substantially lowers the surface energy (Fig.~\ref{fig:ConvexHull_sol}\,a) of Ca substituted in the Mg surface, i.e., solvation promotes solute segregation.

The same energy gain upon solvation is also observed for Al-substituted Mg surfaces, though at a smaller magnitude (Fig.~\ref{fig:ConvexHull_sol}\,b). This energy gain promotes the segregation of Al to the top surface, whereas in vacuum it prefers to reside at the subsurface layers. In this way, the $\theta = \frac{1}{3}$ ordered phase with Al in the top surface layer becomes thermodynamically stable in the presence of water in a small region of the phase diagram (Fig.~\ref{fig:SurfPhaseDiags}d). 

To discuss realistic experimental scenarios we consider a few selected chemical potentials (see Sec.~\ref{sec:chemical_potentials_boundaries}) that represent bulk (blue dashed lines) or solution concentrations in the electrolyte (dashed red and green lines in Fig.~\ref{fig:SurfPhaseDiags}). The low concentrations are consistent with experimental observations~\cite{Nowak}. Using the chemical potentials calculated for the ionic species in water at different concentrations, we observe a significantly  higher thermodynamic stability of Ca$^{2+}$ ions in solution, as compared to Ca in Mg bulk. The higher energetic stability of Ca$^{2+}$ ions in water compared to Ca atoms in Mg bulk favors the dissolution of Ca from the Mg surfaces and prevents the formation of surface structures with higher Ca coverage. A substantially lower propensity to dissolve is found for Al (see red and green dashed lines in Fig. \ref{fig:SurfPhaseDiags}d). Note that the $x$-scale is different for Ca and Al in Fig. \ref{fig:SurfPhaseDiags}c and d. 

\subsubsection{\label{sec:Electrode_nature_of_Al_and_Ca}  Impact of Al and Ca segregation on Mg(0001) in the context of micro galvanic corrosion}

\begin{figure}[t] 
\centering
\includegraphics[width=0.4\textwidth]{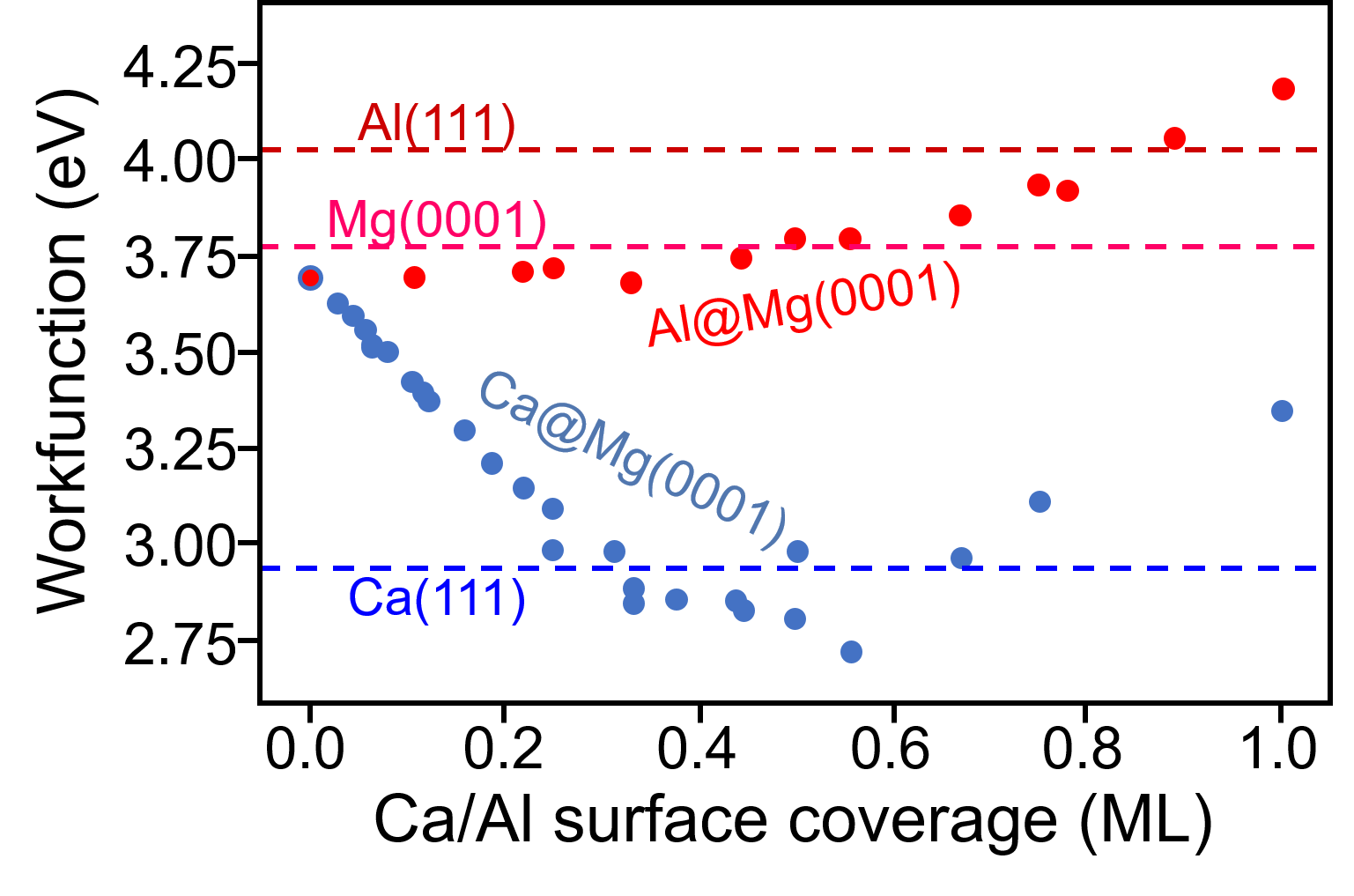}
\caption{\label{fig:Workfct} Impact of Ca/Al substitution on the work function of the Mg(0001) surface. Values for the work functions of the pristine Mg(0001)~\cite{Cheng2015}, Al(111)~\cite{Marzari2009} and Ca(111)~\cite{Wang2014} are marked by dashed lines.}
\end{figure}

The opposite dissolution tendencies of Ca and Al, together with the Al/Ca coverages obtained from the computed surface phase diagrams and the computed electrode potential (work function) of these structures, have direct implications for the corrosion behavior of Mg upon alloying with these elements. To discuss these relations we compute the work functions for the various surface structures and coverages (Fig.~\ref{fig:Workfct}), since the work function indicates the ease for removing electrons from the surface of a material. More details on the calculation are provided in SM, Sec. I~\cite{SM}. For the Mg(0001) surface the work function increases with increasing coverage of Al and shows the opposite trend (i.e. it decreases) for Ca substitution up to ca. 0.6\,ML. From the phase diagrams we know that Al favors high coverages on Mg, while the favorable coverages for Ca lie right within the range of work function decrease. The increase of the work function in the presence of Al means that Al impairs the withdrawal of electrons from Mg, which makes it harder to oxidize the Mg surface. In contrast, the lower work functions for Ca containing surfaces indicates that Ca makes the withdrawal of electrons from the Mg surface easier and therefore promotes its oxidation.

In the context of micro galvanic corrosion these trends in the work functions indicate that Ca is likely to act as anode promoting the electron withdrawal, while Al is likely to act as cathode on the Mg surface, in good agreement with observations of corrosion experiments mentioned in the introduction, i.e., Mg shows anodic potentials in the presence of Ca and cathodic potentials in the presence of Al~\cite{Kirkland2011, Sudholz2011, Leslie2017}. Here, the anodic nature of Ca implies that Ca prefers to oxidize to Ca$^{2+}$ and dissolve from the Mg surface into the electrolyte, whereas Al prefers to remain in the Mg surface. This dissolution propensity and the anodic nature of Ca in Mg surfaces provides support to the experimental observations of higher corrosion rates when the Ca content in Mg alloys is increased~\cite{Kirkland2010, Nabilla2016, Gusieva2015}.

\section{\label{sec:Conclusions} Conclusions}

Based on an unified thermodynamic approach that allows us to treat ordered and disordered surface structures in the same framework, and by using chemical potentials as state variables, we construct phase diagrams for Mg surfaces alloyed with Al/Ca including bulk information in the same diagram. For Ca we find a strong tendency to segregate to the surface, as opposed to Al which shows an anti-segregation behavior, preferring to remain in a more bulk like environment. Generating surface segregation isotherms we predict the surface coverage expected for a given bulk concentration and temperature. Accounting for realistic environmental conditions suggested by experiment we identify the relevant surface coverage regimes. Comparing surface phase diagrams for initial state of pristine as cast alloys, i.e. surfaces in vacuum, with ones in contact with an aqueous environment, we find that the built up of a partial solvation shell around the protruding Ca atoms at the surface leads to the stabilisation of a low coverage ordered structure absent in vacuum. Furthermore we expect, based on the significantly higher thermodynamic stability of Ca$^{2+}$ ions in solution as opposed to Mg bulk, but also to Al$^{3+}$ ions in solution, and the resulting higher dissolution tendency of Ca, that an Al enrichment at the surface  will occur under corrosive conditions. Invoking finally materials properties (like the work function) of all involved elements we rationalise the impact of Al and Ca alloying in Mg in the context of micro-galvanic corrosion and conclude that Ca/Al will be anodic/cathodic in Mg, i.e. the presence of Ca will promote corrosion and Al suppress it.  
\\
\begin{acknowledgments}
We acknowledge funding by the Deutsche Forschungsgemeinschaft (DFG, German Research Foundation) through SFB1394, project no. 409476157.
\end{acknowledgments}


\bibliography{references}

\end{document}